\documentclass[12pt]{article}
\usepackage{graphicx} % Required for inserting images
\usepackage{fullpage}
\usepackage{natbib}
\usepackage{hyperref}
\usepackage{amsmath}
\usepackage{tikz}

\title{Commentary on \cite{Guylletal2023}: Misuse of Statistical Method Results in Highly Biased Interpretation of Forensic Evidence
} 
\author{Michael Rosenblum\footnote{Department of Biostatistics, Johns Hopkins University, Baltimore, MD} \footnote{Corresponding author: Michael Rosenblum  \url{mrosen@jhu.edu}}, Elizabeth T. Chin$^*$, Elizabeth L. Ogburn$^*$, \\ Akihiko Nishimura$^*$, Daniel Westreich\footnote{Department of Epidemiology, UNC-Chapel Hill, Chapel Hill, NC}, Abhirup Datta$^*$, Susan Vanderplas\footnote{Department of Statistics, University of Nebraska Lincoln, Lincoln, NE}, \\Maria Cuellar\footnote{Departments of Criminology; Statistics and Data Science, University of Pennsylvania, Philadelphia, PA}, William C. Thompson\footnote{Departments of Criminology, Law, and Society; Psychology and Social Behavior; and Law, University of California, Irvine, CA.}
}

\begin{document}

\maketitle
\section{Summary}
Since the National Academy of Sciences released their report outlining paths for improving reliability, standards, and policies in the forensic sciences \cite{NAS2009}, there has been heightened interest in evaluating and improving the scientific validity within forensic science disciplines. \cite{Guylletal2023} seek to evaluate the validity of forensic cartridge-case comparisons. However, they make a serious statistical error that leads to highly inflated claims about the probability that a cartridge case from a crime scene was fired from a reference gun, typically a gun found in the possession of a defendant. It is urgent to address this error since these  claims, which are generally biased against defendants, are being presented by the prosecution in an ongoing homicide case where the defendant faces the possibility of a lengthy prison sentence \citep{homicide_case1}.

\section{Error in Statistical Reasoning: Equiprobability Bias} \label{sec_intro}
Firearms examiners try to determine whether a cartridge case from a crime scene was fired from a reference gun.
They do this by comparing surface contour patterns on the crime scene cartridge case to those on cartridge cases fired from the reference gun. The two possible ground truth states are called ``same source" and ``different source", meaning that the crime scene cartridge case was fired from the reference gun or some different gun, respectively.
 A key goal of \citet{Guylletal2023} is to 
 estimate the conditional odds (called  ``posttest odds") that a crime scene cartridge case was fired from a reference gun given the firearms examiner's decision. %We represent this as $\mbox{odds}(\mbox{same source} | \mbox{decision})$. 
 Using Bayes rule, they represent the posttest odds as the product of (i) an assumed \emph{prior} odds (called ``pretest odds") of same vs. different source ground truth, and (ii) the likelihood ratio (LR)\footnote{\cite{Guylletal2023} estimate the LR using data from an experiment to assess how frequently forensic examiners make a particular decision when the ground truth is \emph{same source} or \emph{different source}.  We focus in this manuscript on the pretest odds, but refer the reader to \cite{cuellar2022probabilistic} for information about potential sources of bias in the LR estimate.} of the examiner's decision given same vs. different source. 
% Evidence abounds that to the extent that these data exist they may exhibit systematic bias; we focus this critique on the determination of a pretest odds but refer the reader to *** for more information on the determination of an LR.} of a forensic examiner's decision given same vs. different source ground truth.
 
\iffalse
 \citet{Guylletal2023}  apply a Bayesian updating procedure that begins with an assumed \emph{prior} odds (called the ``pretest odds") on the ground truth (``same source" vs. ``different source") and updates based on the likelihood ratio (LR) of a forensic examiner's decision given same vs. different source ground truth:
    \begin{equation}
      \label{eq}
      \begin{aligned}
        \text{posttest odds}_{same-source} &= \text{pretest odds}_{same-source} \times LR \text{, where}\\        
        odds_{same-source} &= \frac{Pr(\text{same source})}{1-Pr(\text{same source})} \text{ and}\\
        LR_{same-source} &= \frac{Pr(\text{decision}|\text{same source})}{Pr(\text{decision}|\text{different source})}
      \end{aligned}
    \end{equation}

\fi

Next, 
\citet[pp.3-4]{Guylletal2023} 
describe how the above Bayesian procedure can be used by 
``triers of fact in the legal
system, such as judges and juries, who are tasked with evaluating
forensic decisions"; specifically, they state that such triers of fact can set   ``pretest odds equal 1" in (i) above to represent ``a situation corresponding to being initially
unbiased and withholding all judgment as to a comparison’s ground-truth status".\footnote{A related issue not addressed here is whether it is appropriate for forensic scientists (rather than triers of fact such as a judge or jury) to make determinations about prior beliefs/odds. See e.g., \cite{thompson2013role,lund2016likelihood} for discussions of this issue.}
 This claim is incorrect: in fact, such an assumption is substantially \emph{biased}.

By definition, ``pretest odds equal 1" represents the belief that, a priori (i.e., before the firearms examiner's decision is known), it is equally likely that the ground truth is \emph{same source} or \emph{different source}. And while there are indeed two possible ground truths, there is no reason to assume that they are equally likely. This becomes more clear when we consider that the ground truth \emph{same source} requires that a single gun---the reference gun---fired the cartridge case, while the alternative ground truth \emph{different source} requires that the cartridge case was fired by any gun \emph{other} than the reference gun.\footnote{More precisely, {\em different source} means that the crime scene cartridge case was fired by a  ``compatible" gun, i.e.,  one with class characteristics (e.g., the caliber) matching those of the crime scene cartridge case. In other words, the compatible guns are all the guns that are consistent with the gross characteristics of the crime scene cartridge case and are therefore candidates for having fired it.}  
  If there are more than 2 possible guns that could have fired the crime scene cartridge case, then allocating 50\% prior probability to the defendant's gun and the remaining 50\% prior probability to be divided in some manner across all the other guns is, in effect, biasing the pre-test odds toward the reference gun compared to any other individual gun.
%  If there were only 2 guns in the world, then maybe this would make sense. But there are potentially many more candidate guns, and that is the crux of the error. 
The fallacy underlying the above error is the false claim that being unbiased about two possible ground truths (i.e., same, different source) implies that one should believe these ground truths are equally likely (Figure \ref{fig:fig1}). 
This fallacy is an example of what is sometimes called \emph{equiprobability bias} \citep[p.119]{equiprobability_bias_2014}.

\iffalse
\svp{SVP: I think it's worth getting into the different information which can factor into a prior here - prevalence, Blackstone's ratio, etc. I've included such a section at the back of the paper, if you wish to add it. The important part to emphasize is that the prior belongs to the trier of fact, not the statistician or forensic expert. I think also including Figure \ref{fig:fig2} at this point is a great idea... it will break up the paper a bit and make it easier for people to understand the argument. You could even inset figure 1 (or a version of it) showing the scales with varying \#s of firearm pictures... When I generated a version of Fig 2, I used 3.5 million as the total number of guns; this corresponds to the number of Beretta 92s built according to Wikipedia.} 

\mc{I was going to say the same thing: The prior odds should be determined by the trier of fact. In light of this, the fact that the defendant should be considered innocent until proven guilty should at least weigh into the prior odds to be in favor of the defendant. In that case. My paper in JRSSA has a short discussion about this: https://rss.onlinelibrary.wiley.com/doi/10.1111/rssa.12962, uploaded to the Overleaf folder. Should we include Bayes' rule here?}
\fi 

We further illustrate this bias using the following analogy.  %[I WOULD MOVE THE SHARED BIRTHDAY SECTION HERE AND DELETE THAT SECTION HEADER. WOULD ALSO DELETE FIRST SENTENCE OF THAT SECTION]
%\section*{Birthday cake analogy}
%What is an unbiased, fair way to cut a birthday cake? If there are 10 children at a birthday party, one would cut equal sized slices for everyone. But the birthday boy objects: ``there are two groups here: me (special because it is my birthday) and everyone else. Unbiased means each group should get half the cake." This unfairly allocates cake in favor of the birthday boy. 
%\section{Shared Birthday Analogy}Instead of ``same source" or ``different source," consider a version of this question that will be familiar to any reader who has taken a basic probability course. 
Suppose an expert is asked to determine, without knowing anything about your birthday besides that it is one of the 365 calendar days ``compatible" with being a birthday, the probability that you have the same birthday as George Washington. What is an unbiased prior probability for having the same birthday vs. different birthdays? The probability of having the same birthday is approximately $1/365=0.3\%$ and that of having a different birthday is $364/365 \approx 99.7\%$. But the logic of Guyll et al. implies that, instead of considering each possible birth date to be equally likely, an unbiased observer should instead consider the chances of the two scenarios ``same birthday" and ``different birthday" occurring to be equally likely at $50\%$ each. 
This illustrates, intuitively, the problem with the above logic: the number of possible birthdays is critically important but is ignored, just as the number of candidate guns that could have produced the crime scene evidence is ignored.

Unlike birthdays, the number of candidate guns will typically be unknown, which makes determining the pretest odds challenging. Attempting to estimate an ``unbiased" pretest odds is non-trivial, subjective, and likely sensitive to assumptions. It is out of the scope of our expertise to posit realistic but unbiased priors. However, \cite{Guylletal2023}'s assertion that an unbiased probability is universally 50\% for every reference gun %, without consideration of any circumstances specific to the crime, gun, how the gun was identified, etc. 
 should be rejected out of hand.

\section{Impact of Error on Claims about Probative Value of Firearms Examiner Decisions}

Concretely, the erroneous claim of \citet[pp.3-4]{Guylletal2023} of ``being initially
unbiased and withholding all judgment as to a comparison’s ground-truth status"  is equivalent to assuming a 50\% prior probability that the reference gun (and no other gun) fired the crime scene cartridge case.
%, i.e., the odds that  the crime scene cartridge case was fired from the reference gun given the firearm exhttps://www.overleaf.com/project/64cd3d8395879ca43909b8f9aminer's decision. 
Multiplying the prior by the aforementioned LR using Equation 4 of \citet[p.3]{Guylletal2023} propagates this error. In the upper left of Guyll et al.'s Table 3, for example, this approach leads to post-test odds of 177.458:1 (equivalent to 99.4\% probability). 

Dr. Guyll presented the above argument in an ongoing homicide case. Specifically, he asserted that if the firearms examiner made an ``identification" decision (i.e., a ``match"), then an initially unbiased trier of fact should now believe that there is a 99.4\% probability that the crime scene cartridge case was fired from the reference gun \citep[p.67]{homicide_case1}. 
He goes on to state: ``I would consider that to be
extreme [sic] strong support for making the judgment in line with
the forensic decision." \citep[pp.69--70]{homicide_case1}.
The argument is incorrect because it relies on the erroneous claim in the first sentence of this section.  
%In the Supplementary Material, we present a longer excerpt of Dr. Guyll's testimony that includes the above.

Since the posttest odds are highly dependent on the  pretest odds, the error of \citet[pp.3-4]{Guylletal2023} is not innocuous; to the contrary, it can result in highly inflated estimates of the posttest odds, which could lead judges and jurors in criminal trials to grossly misinterpret the forensic evidence.

Consider the case where a firearms examiner's decision is an ``identification". As described above, combining  \cite{Guylletal2023}'s likelihood ratio estimate of $177.458$ with their 50\%/50\% prior on same vs. different source ground truth results in the posterior probability of same source  $99.4$\%. However, if one uses smaller priors, the posterior probability decreases rapidly. For simplicity, consider the case where there are $n$ possible guns and where the prior probability of same source ground truth is set to $1/n$.\footnote{In all of our examples, we use the framework of \emph{equiprobable events} to calculate probabilities simply by enumerating the number of, in this case guns, comprising an event. In this case the event ``same source" includes just one gun and the event ``different source" includes many. One possible refinement would be to move away from the equiprobable events framework and, for example, weight guns as having more prior probability of having fired the bullet if, e.g., they were known to be used in previous similar crimes.} As we increase $n$,  the posterior probability of same source decreases from $99.4$\% (n=2, equivalent to \cite{Guylletal2023}'s prior) to $64.2$\% (n=100) to $15.1$\% (n=1000) to $1.7$\% (n=10,000) (Figure \ref{fig:fig2}).  In an urban area, of course, the number of possible firearms is likely quite large. We are not suggesting to base the prior odds solely on the number of possible firearms, rather using it illustratively to show the sensitivity of the posterior to the choice of prior.

\cite{Guylletal2023} misconstrue the experimental performance of forensic decisions with the probability that a forensic decision reflects the ground truth. In the Abstract, \citet[p.1]{Guylletal2023} state, ``Considering probative value, which is a decision’s usefulness for determining a comparison’s ground-truth state, conclusive decisions predicted their corresponding ground-truth states with near perfection." As we showed  above, the posterior, which \citet[p.4]{Guylletal2023} equates to the probative value, is highly dependent on the prior. \cite{Guylletal2023}'s arbitrary and biased prior odds of 1 (which they also built into their experimental design by assigning same and different source test cases  each with 50\% probability) was used to compute the posttest odds, rendering the latter arbitrary and biased as well.

\iffalse
\textbf{AD: I'm not sure that we give enough justification to claim that `that there is no fair way to calculate prior odds', I think we make the case that it is a challenging task, and one that varies from case to case. Also, there is overlap of this paragraph with the discussion in in Section 5.}
\svp{I've made a couple of edits to this and think  it's a bit better, hopefully?}
%\section{Cancer screening analogy}
\fi

\iffalse

\section{Erroneous Calculation of Posterior Probabilities}
\cite{Guylletal2023}'s assertion erroneously suggests that unbiased triers of fact only need to consider the performance of a forensic technique, as , to determine the posterior probability a firearms examiner's decision reflects the ground-truth state. 
\svp{\cite{Guylletal2023} motivates this claim by establishing the likelihood ratio as free of the need to evaluate the prevalence, and then replacing the prior odds (where the prevalence information should be incorporated) with the biased prior odds of 1.}
This faulty logic yields incorrect estimates of these posterior probabilities %, also commonly known as Positive Predictive value (PPV) and negative predictive value (NPV),
and could lead judges and jurors in criminal trials to grossly misinterpret forensic evidence. Erroneous estimates from this study have already been presented in criminal cases as described at the end of Section~\ref{sec_intro}. \
\fi

\cite{Guylletal2023} acknowledged that the prevalence in casework of same source ground truth (i.e., the marginal probability of same source ground truth) is unknown and therefore one should consider different possible values of it when computing the positive predictive value (posterior probability) of same source ground truth given a firearms examiner's decision. However, this did not stop them from asserting that an unbiased trier of fact  believes equal prior probabilities for same and different source ground truth, which as we argued above is  incorrect. \cite{Guylletal2023}'s prior is not unbiased, rather an assumption that results in an estimated positive predictive value close to the upper bound of 100\%. 
  Given the high reliance of the posterior probability on the choice of prior, we urge careful consideration of the assumptions that underlie priors, resulting uncertainty in the posterior probability, and interpretation of results in the legal context.

\section{Conclusion}

\cite{Guylletal2023} made a serious statistical error that could lead judges and jurors in criminal trials to grossly misunderstand how to interpret forensic evidence. The error should be acknowledged and immediately corrected. 
%In brief, the error is that a uniform prior distribution (i.e., 50\%/50\% prior odds) on same vs. different source ground truths does not represent an unbiased belief about whether the reference gun fired the crime scene cartridge case or not.

\newcommand{\gunVoffset}{0.4}
\newcommand{\gunHoffset}{0.4}

\begin{figure}[h!]
\centering 

\newcommand{\numStories}{1}
\begin{minipage}[c]{.48\linewidth}
\centering
\scalebox{.8}{% Scale the figure by the specified factor
	\begin{tikzpicture}
	\node[inner sep=0pt] (russell) at (0,0)
	    {\includegraphics[width=8cm]{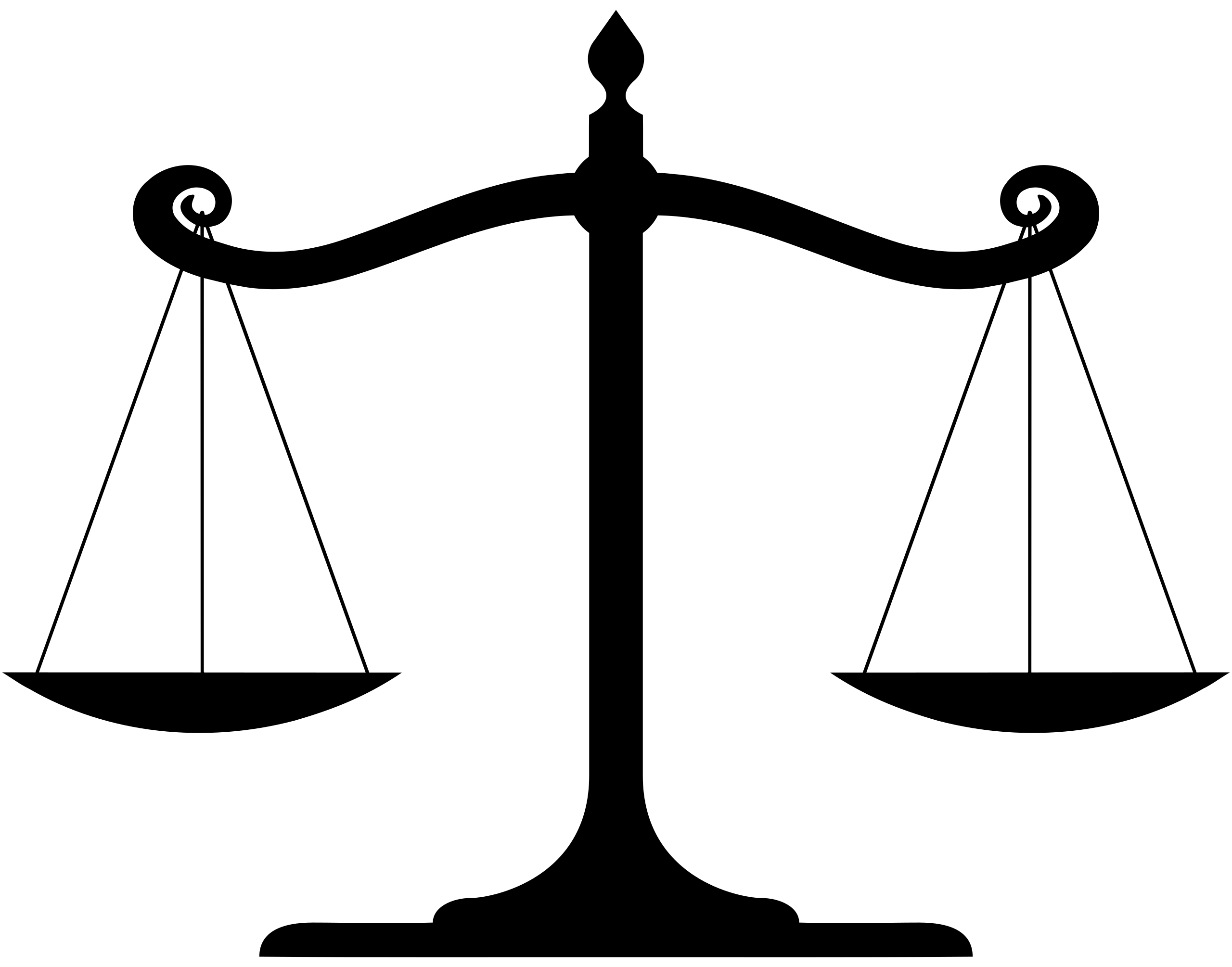}};
	% Gun on the left plate
	\node () at (-2.73, -.89)
		{\includegraphics[width=1cm]{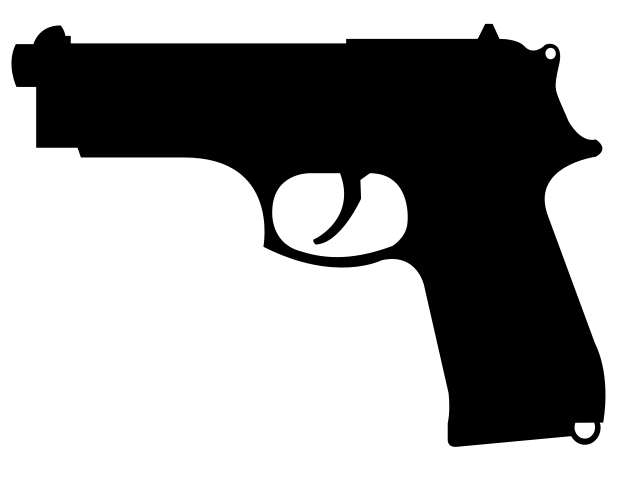}};
	% Gun on the right plate
	\foreach \story in {1,...,\numStories} {
		\pgfmathsetmacro\numCols{\numStories - \story}
		\foreach \col in {0,...,\numCols} {
			\node () at ({2.65 - \gunHoffset * \numCols / 2 + \col * \gunHoffset}, {-.89 + (\story - 1) * \gunVoffset})
				{\includegraphics[width=1cm]{gun}};
		}
	}
	\end{tikzpicture}
}
\end{minipage}

\renewcommand{\numStories}{3}
\begin{minipage}[c]{.48\linewidth}
\begin{tikzpicture}
\scalebox{.8}{% Scale the figure by the specified factor
	\node[inner sep=0pt] (russell) at (0,0)
	    {\includegraphics[width=8cm]{balanced_scale_of_justice}};
	% Gun on the left plate
	\node () at (-2.73, -.89)
		{\includegraphics[width=1cm]{gun}};
	% Gun on the right plate
	\foreach \story in {1,...,\numStories} {
		\pgfmathsetmacro\numCols{\numStories - \story}
		\foreach \col in {0,...,\numCols} {
			\node () at ({2.65 - \gunHoffset * \numCols / 2 + \col * \gunHoffset}, {-.89 + (\story - 1) * \gunVoffset})
				{\includegraphics[width=1cm]{gun}};
		}
	}
}
\end{tikzpicture}
\end{minipage}
~
\renewcommand{\numStories}{5}%
\begin{minipage}[c]{.48\linewidth}
\begin{tikzpicture}
\scalebox{.8}{% Scale the figure by the specified factor
	\node[inner sep=0pt] (russell) at (0,0)
	    {\includegraphics[width=8cm]{balanced_scale_of_justice}};
	% Gun on the left plate
	\node () at (-2.73, -.89)
		{\includegraphics[width=1cm]{gun}};
	% Gun on the right plate
	\foreach \story in {1,...,\numStories} {
		\pgfmathsetmacro\numCols{\numStories - \story}
		\foreach \col in {0,...,\numCols} {
			\node () at ({2.65 - \gunHoffset * \numCols / 2 + \col * \gunHoffset}, {-.89 + (\story - 1) * \gunVoffset})
				{\includegraphics[width=1cm]{gun}};
		}
	}
}
\end{tikzpicture}
\end{minipage}

  \caption{\cite{Guylletal2023} state that an unbiased trier of fact should initially give equal prior weight (probability) to the single reference gun (same source) and to all other guns combined (different source). These should not balance if there are more than 2 guns total, but by Guyll et al.'s erroneous reasoning they would balance.  }
  \label{fig:fig1}
\end{figure}
\clearpage

\begin{figure}[h!]
\includegraphics[width=6.5in]{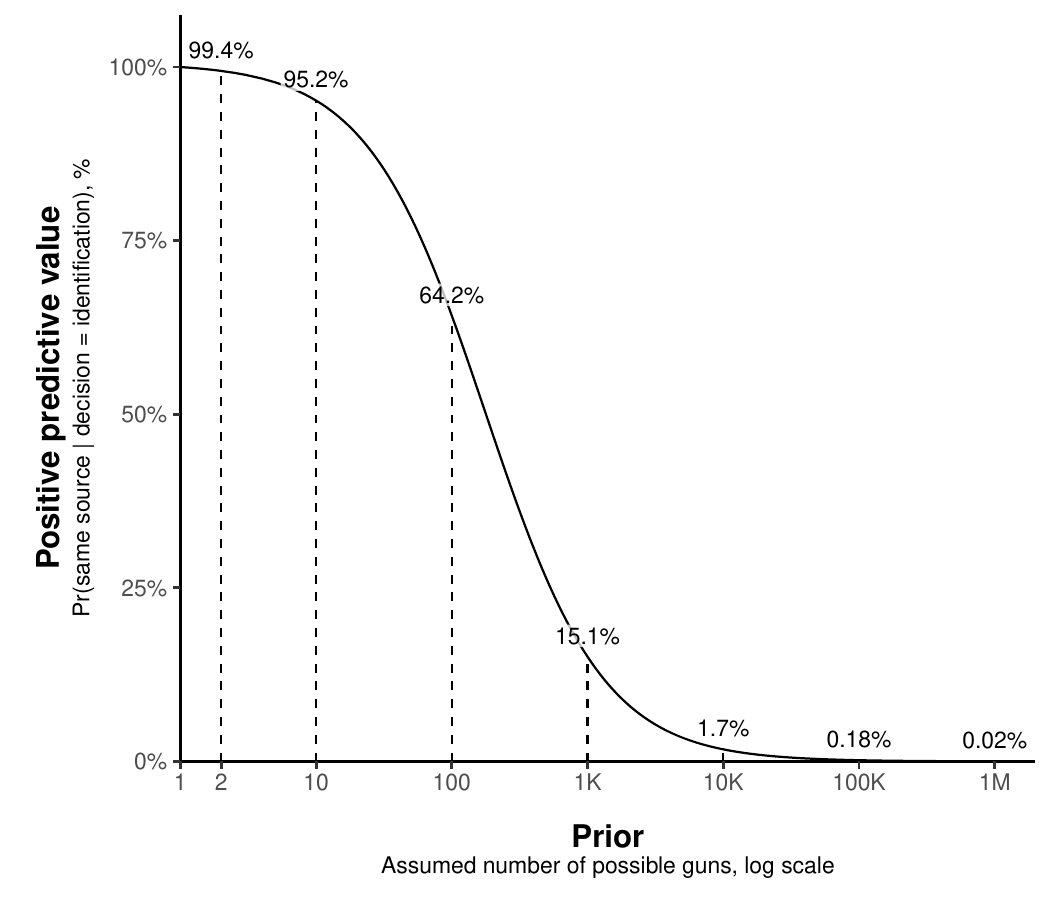}
  \caption{Posterior probabilities are highly dependent on assumptions within the prior. The prior asserted as unbiased in \cite{Guylletal2023} (n=2) corresponds to a positive predictive value that strongly favors assuming a reference gun fired the crime scene cartridge case (same source) given that a firearms examiner's decision of an ``identification".}
  \label{fig:fig2}
\end{figure}

\clearpage

\section{Acknowledgments and Disclosures}
M.R., E.T.C., and E.O. were supported in this research by a Nexus Award  from Johns Hopkins University. 
The opinions expressed herein are those of the authors and do not necessarily reflect the views of The Johns Hopkins University, the D.C. Public Defender Service (PDS), nor anyone else.
We mention the PDS because 
M.R. is an expert witness for it in a homicide case where Dr. Guyll is an expert witness for the prosecution; each is paid for their work on this case, but no such funding was used to support the work on this Comment. 
WT is an expert witness for the Innocence Project, which is involved in the same case as \textit{Amicus Curiae}. 
D.W. and A.D. report no conflicts and had no funding support for this work.
We thank Dr. Charles Poole of UNC-Chapel Hill for his helpful input.
\bibliography{references}
\bibliographystyle{Chicago}

\end{document}